\documentclass{cfdsc}
\usepackage[T1]{fontenc}
\usepackage[latin1]{inputenc}
\usepackage{setspace}
\usepackage{amssymb}

\usepackage{multicol}
\usepackage{mathtools}
\usepackage{epsfig}
\usepackage{float}
\usepackage{lipsum}
\usepackage{lineno,hyperref}
\modulolinenumbers[5]
\usepackage{times}
\usepackage{amsmath}
\usepackage{amsmath,bm}
\usepackage{amssymb}
\usepackage{setspace}
\usepackage{graphics}
\usepackage{fancyhdr}
\usepackage{rotating}
\usepackage{color}
\usepackage{amsmath}
\usepackage{titlesec}
\titleformat{\paragraph}
{\normalfont\normalsize}{\theparagraph}{1em}{}
\titlespacing*{\paragraph}
{0pt}{3.25ex plus 1ex minus .2ex}{1.5ex plus .2ex}
\titleformat*{\subsubsection}{\normalfont}
\usepackage{multirow}
\usepackage{placeins}
\usepackage[table,xcdraw]{xcolor}
\usepackage[square,numbers]{natbib}
\graphicspath{{./plots/}}

\begin{document}

\title{Dynamic modelling of near-surface turbulence in large eddy simulation of wind farms}

\author{Jagdeep Singh\inst{1} \and Jahrul Alam\inst{1}}

\institute{Department of Mathematics and Statistics, Memorial University of Newfoundland, \\ 230 Elizabeth Ave,Newfoundland, NL, A1C 5S7, Canada 
}
\email{jagdeeps@mun.ca}
\email{alamj@mun.ca}

\begingroup
\onecolumn{
{\begin{flushright} \small{{Proceedings of the 29th Annual Conference of the Computational Fluid Dynamics Society of Canada}\\{CFDSC2021}\\{July 27-29, 2021, Online}}
\end{flushright} }}

\maketitle

\endgroup


\begin{multicols}{2}
%
\section{Abstract}
In large eddy simulation of atmospheric boundary layer flows over wind farms, wall-layer models are generally imposed for the surface fluxes without considering the spatial variability of the surface roughness. In this study, we consider the near-surface model in conjunction with square of the velocity gradient tensor to model the adaptive dissipation of turbulence production. The surface roughness is incorporated through Monin-Obhukhov similarity theory for the computational cells immediately adjacent to the Earth's surface. The underlying proposed near-surface model captures the significant amount of Reynolds stresses in the near-surface and is able to maintain the log-law profile in wind farms. The present study indicates that the suggested `near-surface model' is relatively robust in comparison to the classical `near-wall model'. 
\section{Introduction}
  Effects of Coriolis force, atmospheric stability, and meteorological inflow conditions have been extensively studied \cite{calaf2010large,roy2011simulating, meneveau2012top, churchfield2012numerical,wu2013simulation,xie2017numerical} for wind farms. Generally, equilibrium-contingent models are considered for the surface fluxes ignoring the local variations of the surface roughness. For instance, refs\cite{churchfield2012numerical, meneveau2012top} indicates the potential role of local variation of roughness on the performance of wind turbines. In atmospheric boundary layer flows (ABL) over complex terrain, the aerodynamic roughness length ($z_0$) depends on both the height and distribution of various roughness elements \cite{garratt1994atmospheric}. Local variation of the aerodynamic roughness would have potential influence on the inflow turbulence intensity, thereby, effecting the structural response and power production in the large wind farms \cite{churchfield2012numerical}.
  \subsection{Literature Review}
 Typical grid resolutions of large eddy simulation (LES) are insufficient for capturing the dynamic responses of the local variation of roughness of complex terrain \cite{lundquist2012immersed,bao2018large}. The roughness effect of horizontally homogeneous landscape, for example, grass may be treated with a standard wall-modelling approach \cite{stoll2020large}. Such wall-modelling techniques are broadly classified into two categories: In the first category, wall-stress models which corrects the velocity gradient at walls through an algebraic relationship between the wall velocity and the velocity at some distance from the wall. The second category combines LES with Reynolds averaged Navier-Stoke's (RANS) equation. In such a hybrid RANS/LES approach, the subgrid model is switched from LES to RANS in close proximity to the wall. The detailed review of both the techniques were done in several studies \cite[e.g.][]{piomelli2002wall,cabot2000approximate,senocak2007study, piomelli2008wall}.
  In literature, there is a lack of sufficient attempts of understanding how to incorporate the effects of local variations of roughness into the wall-stress or hybrid RANS/LES models. Wall-stress models based on Monin-Obhukhov similarity theory often differs in atmospheric boundary layer flows \cite{moeng1984large,moeng1994comparison,porte2000scale,brasseur2010designing,sullivan2011effect} as compared to the engineering applications \cite{versteeg1995computational, senocak2007study}. In ABL, the log-law is used directly to calculate the local shear stress and imposed an average shear stress computed by LES.
  This technique have been applied in numerical studies of wind-farms \cite{calaf2010large,churchfield2012numerical,stevens2014concurrent}.
  
In this article, we are interested in modelling the transient effect of surface roughness on the near-surface layer of ABL in wind farms. With all the development of wall-layer models for LES, there are still some remaining challenges. First, classification of surface roughness depends on the size and distribution of the roughness elements \cite{wieringa1992updating}. For surfaces containing more than one categories of roughness elements, considering the constant surface flux may not be a good approximation. A complete description of the surface roughness can be found in \cite{stull2000meteorology,garratt1994atmospheric}. Second, a mountainous terrain may contain various roughness elements and it may be complicated to smoothly blend LES with RANS if hybrid RANS/LES model is to be used. Third, such wall-models perform poorly in the presence of relatively large scale obstacles, for example, buildings, topography undulations, and vegetative canopies\cite{stoll2020large}. Therefore, the shortcomings of LES for ABL
flows are often attributed to poor resolution and inadequacy
of subgrid model [7].
\subsection{Present Work}
Wind-farm operates in the surface-layer ($10-20~\%$) of the atmospheric boundary layer which accounts for the major portion of turbulent kinetic energy production and dissipation. The flow in wind-farm is considered to be highly turbulent due to spatial variability of surface roughness and interaction of ABL with wind turbines. As the Earth's surface is approached, the length and time scales considerably decreases and turbulence anisotropy increases. This phenomena is partly resolved by the LES. The vortex-tubes in the surface-layer as well as in  close proximity to the surface, are stretched by the principal rate of strain. In the presence of large array of wind turbines, the coherent structures in the surface-layer will also be influenced by wake-vortex interactions. We  follow the modern view of Townsend's attached hypothesis \cite{marusic2019attached, townsend1980structure} that momentum-carrying eddies near the Earth's surface are governed by the mean momentum-flux and mean-shear without any explicit dependence on the normal distance to the surface \cite{marusic2019attached}. \cite{chung2009large} numerically observed that the stretched vortex-tubes near a surface governs the majority of the wall-shear stress.

In Smagorinsky subgrid scale model, second invariant of the velocity gradient $\mathcal{S}_{ij}$ is considered. In contrast, we are considering the square of the velocity gradient tensor \cite{nicoud1999subgrid,alam2018large,bhuiyan2020scale} $\mathcal{G}_{ij}=(\partial u_{i}/\partial x_{k}) (\partial u_{k}/\partial x_{i})$ for the development of near-surface model for the atmospheric turbulence in wind-farms in which eddy viscosity is dynamically computed using rate of strain and rate of rotation. We aim to study a dynamic approach to model surface exchange through dynamic calculation of shear-stress at every grid point adjacent to the Earth's surface. We propose a methodology which adjusts the eddy viscosity dynamically using second invariant of the square of the velocity gradient tensor, following an additional adjustment of the  eddy viscosity arising from the shear stresses due to the surface roughness.

In section 3, we discuss the formulation and implementation of the square of the velocity gradient tensor, near-surface flux and wind-farm modelling. In section 4, numerical assessment of the proposed `near-surface' model is discussed in case of turbulent channel flow, single wind turbine and utility scale wind-farm. Finally, in section 5, we summarizes the findings of the present article. 

\section{Methodology}
\subsection{LES of ABL over wind farms}
A large eddy simulation of neutrally stratified ABL flow over wind farms is presented through this work \cite{porte2000scale, calaf2010large, churchfield2012numerical, xie2017numerical}. Atmospheric turbulence around wind farms is characterized by the length scales of the order of $100$~m, which may be resolved by the LES approach using a grid space of $\Delta \approx \mathcal{O}(10~m)$. In LES, filtered component of $\bar{u}_{i}$ is computed by solving the Eq. (\ref{ge1} -- \ref{ge2}). When the mesh is sufficiently fine, prediction of turbulence in wind farm is relatively insensitive to the choice of subgrid models for the subfilter scale stresses\cite{sarlak2015role}. The filtered momentum equations,
\begin{equation}
	\frac{\partial \bar{u}_{i}}{\partial t} +\bar{u_{j}}\frac{\partial \bar{u}_{i}}{\partial x_{j}} = -\frac{\partial \bar{p}}{\partial x_{i}}-\frac{\partial \tau_{ij}}{\partial x_{j}} + f_{i},
	\label{ge1}
\end{equation}
and the incompressibility condition,
\begin{equation}
	\frac{\partial \bar{u}_{i}}{\partial x_{i}} = 0,
	\label{ge2}
\end{equation}
are solved numerically, subject to appropriate boundary conditions. Here, $\bar{u_{i}} \big(i = 1, 2, 3\big)$ are the filtered velocity component. The corresponding wave number is $k_{c} = 2\pi/\Delta$, where $\Delta = \left(\Delta x \Delta y \Delta z \right)^{1/3}$. The sub-filter scale stress tensor in Eq. (\ref{ge1}) is defined as 
$\tau_{ij} =\overline{u_{i}u_{j}} - \bar{u_{i}}\bar{u_{j}}$, and its deviatoric part is represented through the classical Smagorinsky model. i.e. $\tau_{ij}^{d} = - 2\nu_{sgs}  \bar {\mathcal{ S}_{ij}}$.
Assuming that the unresolved flow is isotropic, the eddy viscosity is usually given by
\begin{equation}
	\nu_{sgs} = C_{s}\Delta_{les}^{2}\sqrt{2\bar{\mathcal{S}_{ij}}\bar{\mathcal{S}_{ij}}},
	\label{nusgs3}
\end{equation}  
where $\bar{\mathcal{S}_{ij}} = 1/2\left(\partial \bar u_{i}/\partial x_{j} + \partial \bar u_{j}/\partial x_{i}\right)$ is the symmetric part of the velocity gradient tensor. An estimate for the Smagorinsky constant is $C_{s} \approx 0.17$.
In ABL flows, the scale of energetic eddies decreases near $z = 0$, where $z=x_{3}$, is the surface-normal direction. One way to adjust eddy
viscosity to the rate of dissipation of near-surface turbulence is to adjust the filter width as $\Delta_{les}^{-2} = \left[C_{s}\Delta\right]^{-2} + \left[k\big(z+z_{0}\big)\right]^{-2}$, 
where $k = 0.41$ is the Von Karman constant, and $z_{0}$ is the roughness
length. This is a numerical approach that squeezes the spectrum of the resolved flow in the near-surface region. Another method that is commonly employed by the atmospheric sciences research
community is the Deardorff model based on turbulent kinetic energy (TKE) \cite{deardorff1972numerical}. It is necessary to solve the energy transport equation in order to find the subgrid scale TKE, $k_{sgs}$. If the production of turbulence
near $z = 0$ were locally balanced by the rate of dissipation, then $k_{sgs}$ would be adjusted dynamically
as the energetic length scales diminishes near $z = 0$. Under these assumptions, the eddy viscosity is
defined by
\begin{equation}
	\nu_{sgs} = C_{s}\Delta_{les}k_{sgs}^{1/2}.
	\label{nusgs4}
\end{equation}
The models discussed above are designed to account for the flow physics of the atmospheric boundary layer. These schemes do not directly account for the vortex stretching as well as the flow physics around wind turbines\cite{calaf2010large}. Surface layer eddies as well as the tip-vortices are stretched
by the wind shear. Moreover, interactions among multiple wakes in a large wind farm will enhance the rate of turbulence mixing \cite{meneveau2012top}. Numerical studies of turbulence, however, indicate that vortex stretching \cite{Carbone2020, nicoud1999subgrid} as well as the fluctuations of the strain field \cite{leveque2007shear,sullivan1994subgrid} play a dominant role on the average cascade of TKE from the largest to the smallest scales. Numerical studies reported evidences that strain and vorticity fields of the smallest resolved turbulence fluctuations may be privileged in modelling subgrid scale turbulence \cite{alam2018large,bhuiyan2020scale,nicoud1999subgrid}. Let us consider the traceless symmetric part of tensor $\mathcal{G}_{ij} = \left(\partial \bar{u_{i}}/{\partial x_k}\right)\left(\partial \bar{u_{k}}/{\partial x_j}\right)$ given by $\mathcal{G}_{ij}^{d} = \big(1/2\big) \left[\mathcal{G}_{ij} + \mathcal{G}_{ji}\right]$-$\big(1/3\big)\mathcal{G}_{kk}\delta_{ij}$. It can be shown that $ \mathcal{G}_{ij}^{d}\mathcal{G}_{ij}^{d}=\big(1/2\big)\lvert{\mathcal{S\omega}} \rvert + \big(2/3\big)\mathcal{Q}^{2}$, where $\mathcal{S\omega}$ is the vortex stretching vector, $\mathcal{Q} = \big(1/2\big)\big(\mathcal{R}_{ij}\mathcal{R}_{ij}-\mathcal{S}_{ij}\mathcal{S}_{ij}\big)$, and $\mathcal{R}_{ij}+\mathcal{S}_{ij}=2\left(\partial \bar{u_{i}}/{\partial x_j}\right)$. Thus, the second invariant of the tensor $\mathcal{G}_{ij}^{d}$ detects strain, rotation, and stretching of vortices. Using a simple dimensional analysis, we express the subgrid-scale TKE dynamically as
\begin{equation}
	k_{\hbox{\tiny sgs}} = \frac {\Delta_{les}^{2}\left(\big(1/2\big) \mid \mathcal{S\omega} \mid + \big(2/3\big)\mathcal{Q}^{2}\right)^{3}} {\left[\left(\mathcal S_{ij} \mathcal S_{ij} \right)^{5/2} + \left(\big(1/2\big) \mid \mathcal{S\omega} \mid + \big(2/3\big)\mathcal{Q}^{2}\right)^{5/4}\right]^{2}}
	\label{ksgs5},
\end{equation}
Substituting the expression for $k_{sgs}$ given by Eq (\ref{ksgs5}) into the Eq (\ref{nusgs4}), we obtain the eddy viscosity $\nu_{sgs}$
that is required to model the deviatoric part of the stress tensor $\tau_{ij}$ appeared in Eq (\ref{ge1}). To arrive at
Eq (\ref{ksgs5}), we compare the expression of the eddy viscosity given in \cite{deardorff1972numerical} with that given
in \cite{nicoud1999subgrid}. The quantity $(\mathcal{G}_{ij}^{d}\mathcal{G}_{ij}^{d})^{3/2}$ varies asymptotically leading to $\nu_{sgs} = \mathcal{O}(z^{3})$
as $z \to 0$. This property of $\mathcal{G}_{ij}^{d}$ is useful for LES of atmospheric turbulence around wind farms, where the viscous layer is often not resolved. The size of the first grid cell adjacent to $z = 0$ is larger than the depth of the viscous sublayer. The eddy viscosity, given by Eq (\ref{nusgs3}-\ref{ksgs5}), is adjusted dynamically in local
regions of vortex stretching and strain self-amplification. To evaluate $C_{w}$, let us assume that the average rate of dissipation provided by the subgrid model Eq (\ref{nusgs3}-\ref{ksgs5}) is about the same as what would be obtained from the classical Smagorinsky model. Comparing Eq (4) with the expression for the eddy viscosity given by Eq (\ref{nusgs3}), we get
\begin{equation}
	C_{w} = C_{s}^2 \frac{\sqrt{2\mathcal{S}_{ij}\mathcal{S}_{ij}}}{k_{sgs}^{1/2}},
\end{equation}
where $k_{sgs}$ is given by Eq (\ref{ksgs5}) and $C_{s} = \left(1/\pi\right) \left(3C_{k}/2\right)^{-3/4}$.
%
\subsection{Near-surface turbulence and wind-farm modelling}
\subsubsection{Near-surface flux modelling}
\begin{figure}[H]
\centering{\psfig{figure=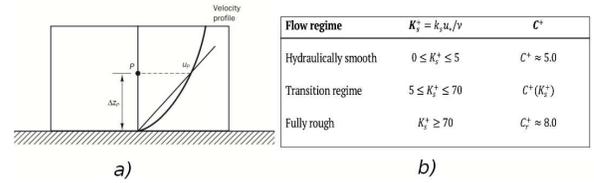,width=0.9\columnwidth}}
\caption{a) Plot shows the first grid cell from the Earth's surface, where, $\Delta z_{p}$ is the variation of cell center height from the surface, $u_{p}$ is the velocity at the cell center. b) Flow regimes based on dimensionless characteristic number ($K_{s}^{+}$) \cite{schlichting2016boundary}.}
	\label{profile}
\end{figure}

In the present work, we consider a situation where the spatial variability of terrain becomes too irregular to be captured by the immersed boundary method \cite{bao2018large} or terrain following method. Consider an equilibrium-contingent wall-stress model based on Monin-Obhukhov similarity theory for the smooth surface, the velocity profile may be assumed to satisfy a logarithmic law \cite{deardorff1970,versteeg1995computational,piomelli2002wall,piomelli2008wall,bose2018wall},
\begin{equation}
	u(z^{+}) = u_{*}\left[\frac{1}{k} \ln (z^{+}) + C^{+}\right].
	\label{loglaw}
\end{equation}
 where $C^{+}$ is a constant of integration, $z^{+}=zu_{*}/\nu$, is a dimensionless distance, and $u(z^{+})$ is the stream wise velocity averaged over a horizontal surface. Eq. (\ref{loglaw}) can be modified to account for the surface-roughness and Fig \ref{profile}b shows that flow regimes can be categorized into three categories based on dimensionless characteristic number, $K_{s}^{+}=k_{s}u_{*}/\nu$, obtained by scaling the sand-roughness height $k_{s}$ with viscous length scale $\delta_{v}$ \cite{nikuradse1950laws, schlichting2016boundary}. This approach may be suitable in the context of the wake-enhanced top-down model of wind farms \cite{frandsen2006analytical} where one assumes a roughness sub-layer that is characterized by a constant stress and zero pressure gradient (\cite[see][]{calaf2010large,stevens2014large}).
\begin{equation}
	\tau = \nu_{sgs} \frac{u_{p}}{\Delta z_{p}}. 
\end{equation}
Wall-stress models based on Eq. (\ref{loglaw}) assumes a constant value of friction velocity ($u_{*}$) on the boundary at $z=0$, and  a boundary condition for $\tau_{ij}$ is obtained using Eq.(\ref{loglaw}). As it can be seen from Fig \ref{profile}a that this approach is accurate if local variation of wall-shear stress ($\tau_{w}$) at $z=0$ is not important.
Instead  of predicting the land surface fluxes from equilibrium conditions based on the Monin-Obukhov similarity theory \cite[e.g.][]{brasseur2010designing, arthur2019using,moeng1984large,stoll2020large,wurps2020grid},
we consider the local value of $u_{p}$ at every cell adjacent to the surface which provides a local velocity gradient. Eq. (\ref{loglaw}) provides a local value of friction velocity ($u_{*}$) corresponding to each value of $u_{p}$. Considering $u_{*}^{2} = \tau_{w} $ and $\tau_{ij} = \nu_{sgs}\mathcal{ S}_{ij}$, at $z=p$, we correct the value of eddy viscosity ($\nu_{sgs}$). From Fig \ref{profile}a, it can be clearly seen that if the resolution is not sufficient to capture the gradient, adjusting an eddy viscosity ($\nu_{sgs}$) will correct the shear stress and maintain the log-law. In this approach, one approximates the deviation from the average shear stress,$ \tau_{w} = u_{*}^{2}$, as a function of the instantaneous streamwise velocity component, using this formulation, one arrives at the classical wall stress formulation based on the Monin-Obukhov similarity theory; i.e $\tau_{w}\left(x,y,\Delta z_{p},t\right) = c_{d} u_{p}{u}\left(x,y,\Delta z_{p},t\right)$,where the drag coefficient $c_{d} = \left[k/\ln \left(z_{1}/z_{0}\right)\right]^{2}$ and the subgrid scale eddy viscosity $\nu_{sgs}$ is adjusted in all grid points adjacent to $z = 0$.
\subsubsection{Wind farm modelling}
Simulating the details of atmospheric boundary layer flows (ABL) in wind-farms is prohibitively expensive. For example, approximately $30$ grid points per actuator line are required to capture the tip-vortices accurately around an individual wind turbine~\cite{sarlak2015role}. A comparative study of different parameterization techniques of wind turbine was done by \cite{stevens2018comparison} and they reported that actuator disk can accurately represents the wind turbine if one is interested in the main flow structures. In the present study, we consider an actuator disk model where each wind turbine is represented as a local momentum deficit, and thrust force experienced by each turbine is, $f_{t} = \frac{1}{2} \rho c_{t} A \langle \lvert u_{d} \rvert \rangle^{2}$,
where $ \lvert u_{d} \rvert =\big(1/A \big) \iint \limits \lvert \bar{\textbf{u}} \rvert dy dz$ is the average of the instantaneous velocity $\textbf{u}(x,y,z,t)$ experienced by each rotor, $A = \big(\pi/4\big)D^{2}$ is the area of the rotor. The thrust coefficient of a wake-affected wind turbine is calculated as $c_{t} = C_{T}/\big(1-a\big)^{2}$, where $C_{T}$ is the thrust coefficient based on the momentum theory \cite{stevens2018comparison} and $a$ is the axial induction factor which relates the free stream velocity ($u_\infty$) and the velocity at the rotor disk through $u_{\infty} = u_{d}/(1-a).$

\begin{figure}[H]
 	\centering{\psfig{figure=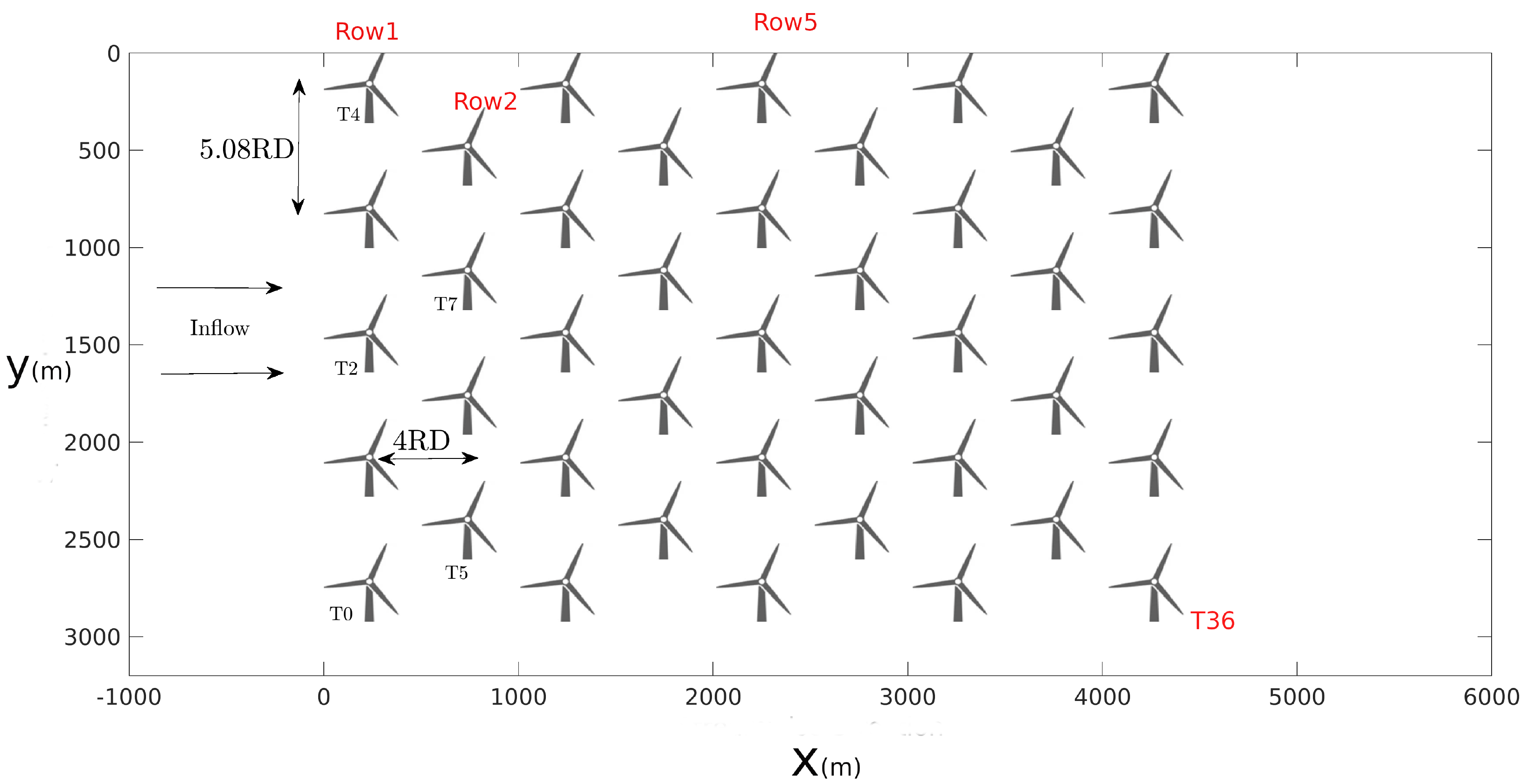,width=1.0\columnwidth}}
 	\caption{2D layout of wind farm used for LES simulation, where spanwise and streamwise spacing between turbines are 5.08 and 8 $\mathcal{R}_{d}$ respectively. $\mathcal{R}_{d}$ of wind turbine used in this study is $126$~m}
 	\label{layout}
 \end{figure}

In the current article, we consider an LES study of atmospheric turbulent flow past an array of 41 wind turbine with the staggered arrangement as shown in Fig \ref{layout}. The rotor diameter and the hub height of each turbine is $\mathcal{R}_{d} = 126$~m and $\mathcal{H}_{hub} = 90$, which represents the \emph{REpower} 5-MW turbine \cite{xie2017numerical}. The computational domain $\left[7048 \times 2880 \times 640\right]~m^{3}$ is discretized into $624\times 256\times 84$ finite volume cells, which is stretched in the vertical direction, thereby, leading to $\Delta z_{min} = 2.29$~m near the bottom boundary, $z=0$, and $\Delta z_{max} = 11.5 $~m near the top boundary ($z=H$). There are 11 grid cells across the rotor in the stream-wise direction.   

\section{Results and discussions}
\subsection{Turbulent channel flow}
\subsubsection{Direct numerical simulation (DNS) and LES}
\begin{figure}[H]
	\centering{\psfig{figure=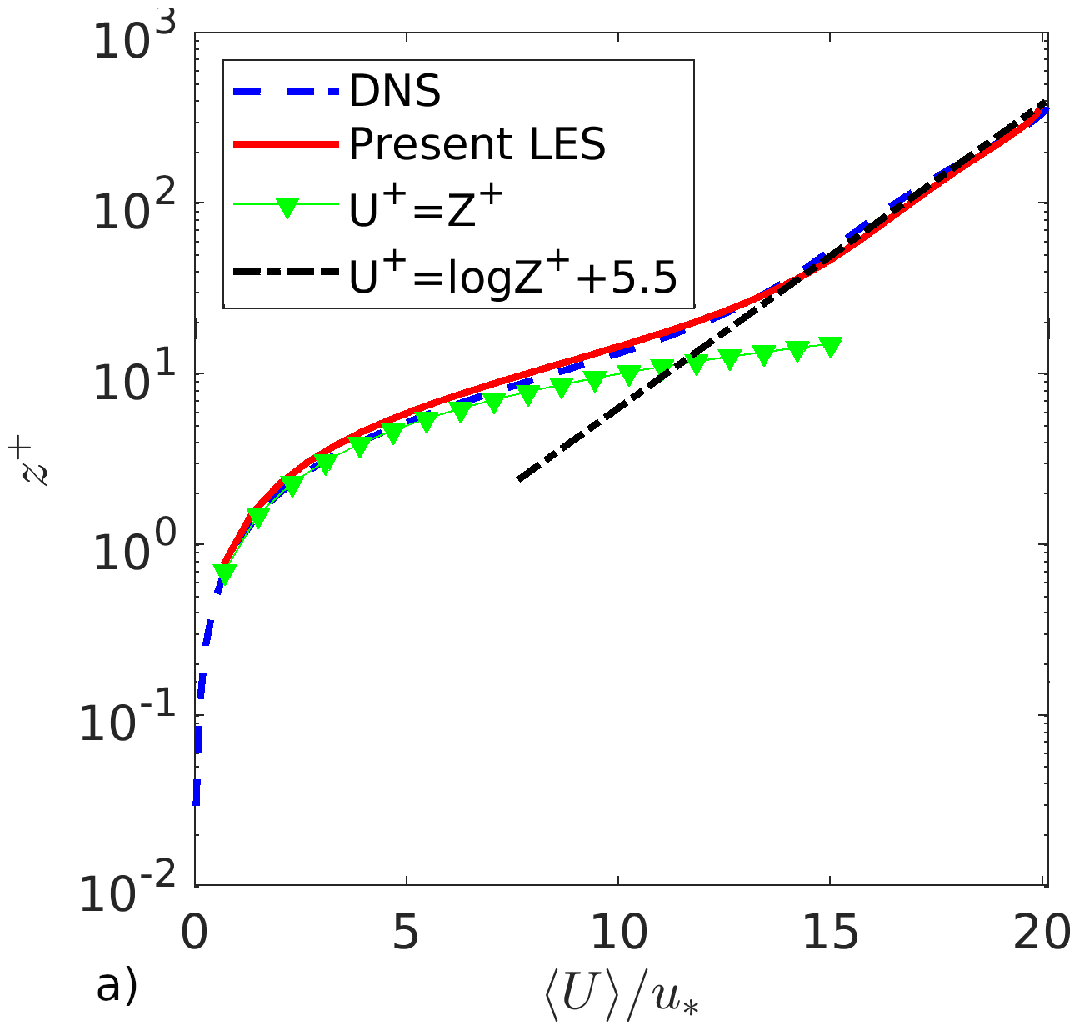,width=0.4\columnwidth}}
	\centering{\psfig{figure=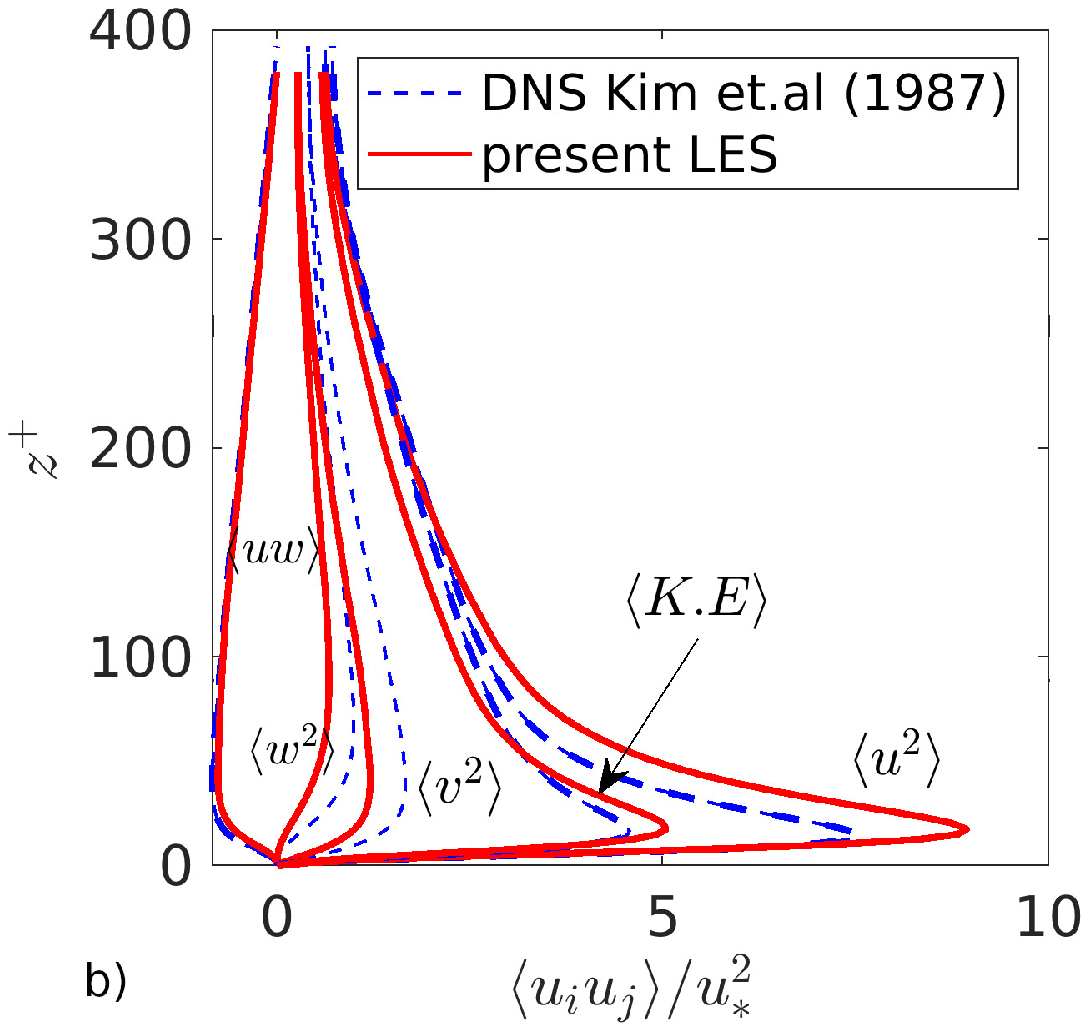,width=0.4\columnwidth}}
	\caption{Plot(a-b) first and second order statistics of fully resolved surface-layer in a fully developed turbulent channel flow at $Re_{\tau} = 395$}
	\label{tc1}
\end{figure}

One can show that \cite[e.g][]{pope2001turbulent, davidson2015turbulence}, the Reynolds stress of near-surface turbulence depends on local vorticity fluctuations $\omega'$; i.e.
$$
\frac{\partial}{\partial z}\Big[\frac{\tau_{13}^{R}}{\rho}\Big] = \Big[\overline{u' \times \omega'}\Big]_{x}.
$$
In LES, we cannot resolve the vorticity fluctuations near a rough or smooth surface, whereas we expect that the subgrid model accounts for the near-surface fluctuations induced by the remote eddies \cite[][]{pope2001turbulent,davidson2015turbulence}. DNS results for turbulent flows in  channels presumably
indicate that fluctuations in both the velocity and the vorticity have been adequately captured. Such fluctuations are not resolved by LES. Here, we consider LES of turbulent channel flows using a grid
$192 \times 96 \times 96$ which is stretched in the vertical direction, thereby, leading to $\Delta z_{min} = 5.2\times10^{-3}$~m at the bottom boundary, $z=0$. At a moderate Reynolds number, $Re_{b}$ = $13,950$, transitions to the appropriate turbulent regime occur naturally in channel flows \cite[][]{pope2001turbulent}. First, we compare the results of LES with that of DNS for a turbulent channel flow \cite[][]{kim1987turbulence}. Next, we discuss LES of channel flows on a coarser grid $96 \times 48 \times 48$, where the near-wall resolution is also coarsened by a factor of 5 ($\Delta z_{min}=2.8\times10^{-2}$~m) with respect to the previous simulation on a finer grid $192 \times 96 \times 96$. For the coarser resolution, we consider that the channel walls are rough and the values of the Reynolds number $Re_{\tau}$ are $395$, $590$, $1000$, $2200$, and $5200$, as shown in Fig \ref{lesdns}(a--b). We demonstrate the comparison results to indicate that the present LES bridges the bulk of the channel, dominated by the fluctuations of the strain and vorticiy, and the proximity of surfaces with prevailed mean shear. Fig \ref{tc1}a displays the profile of the streamwise mean velocity $u^{+} = \langle \overline{u} \rangle/u_{*}$ as a function of the wall normal distance $z^{+}$. The velocity $u^{+}$ is normalized by the wall friction velocity $u_{*}$ , where the wall-normal distance $z^{+}$ is normalized by the length scale $\delta_{\nu} = \nu/u_{*}$ of viscous dissipation. The mean velocity profile exhibits an agreement with the DNS reference data obtained from \cite[][]{kim1987turbulence}. In the near-wall region $z^{+} \leq 10$, the mean velocity varies like $u^{+} = z^{+}$. In the overlap region, $30 $$\leq$$ z^{+} $$\leq$ $300$, the LES data is in a good agreement with the DNS data as well as with the logarithmic profile of the velocity. Fig \ref{tc1}b indicates a slight over prediction of the Reynolds stresses, which may be attributed to the relatively less dissipative numerical scheme, as well as the forcing method, considered in the present work. Nevertheless, the peak value of K.E. ($\approx$ $5$) for the LES data has a better agreement with the value reported by \cite{townsend1980structure,schlichting2016boundary}.
\subsubsection{Influence of surface roughness}
\begin{figure}[H]
	\centering{\psfig{figure=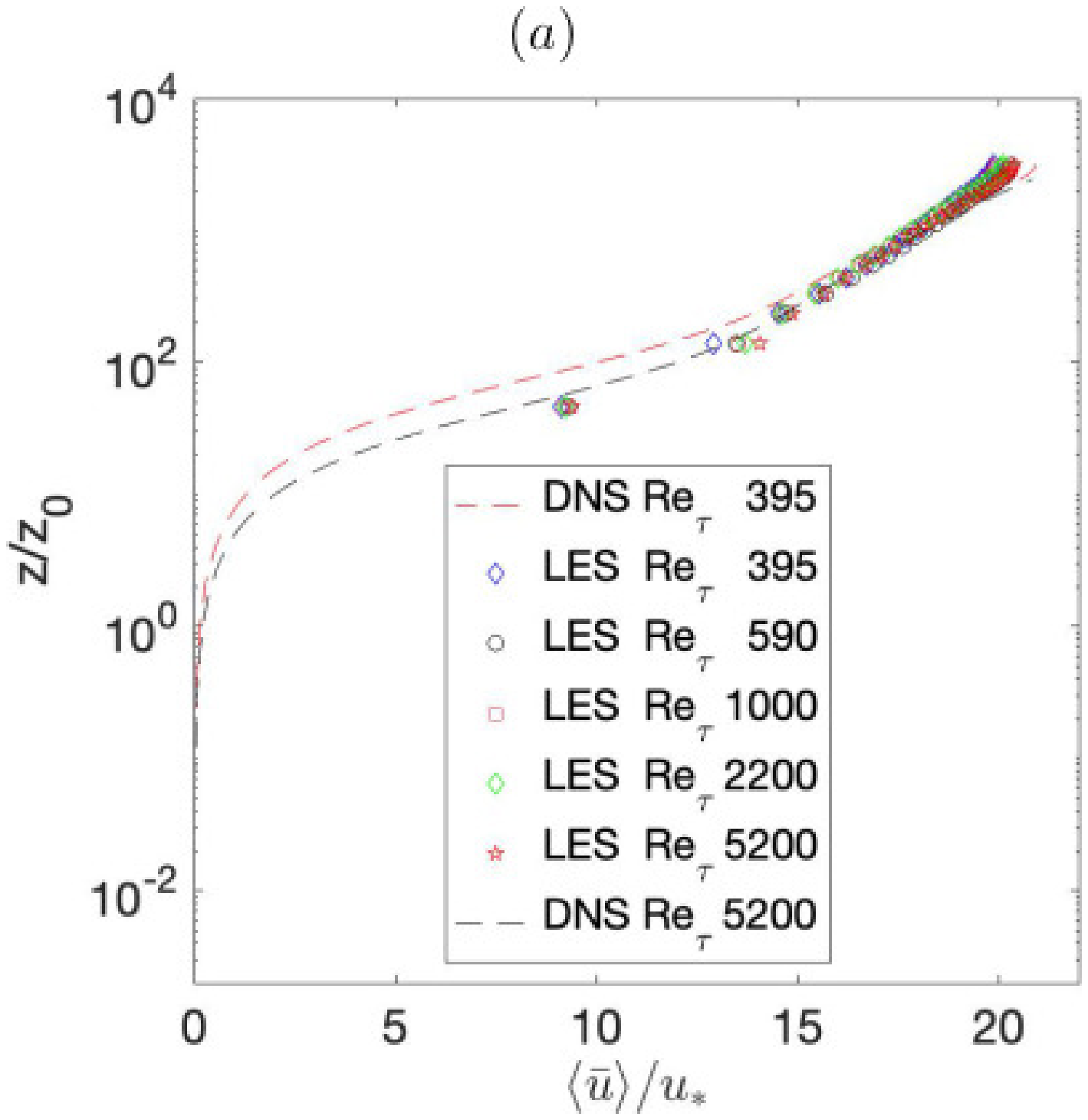,width=0.4\columnwidth}}
	\centering{\psfig{figure=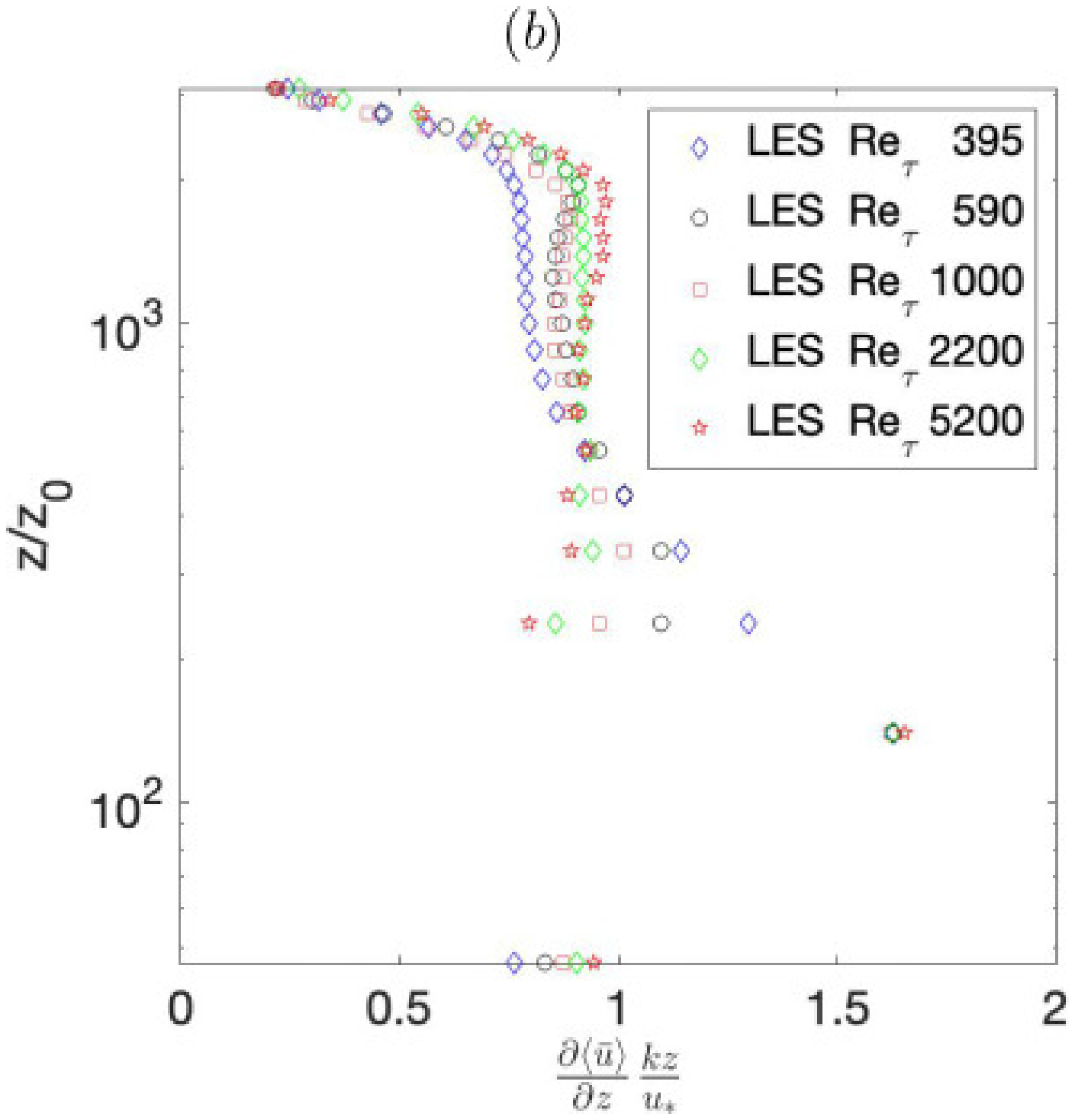,width=0.4\columnwidth}}
	\caption{A comparison of the streamwise mean velocity obtained from present LES study with previously reported DNS studies; (a) mean velocity and (b) the wall-normal gradient of the velocity. The values of Reynolds number are $\mathcal{R}e_{\tau} = 395,\,590,\,1000,\,2200~\&~5200$.}
	\label{lesdns}
\end{figure}

According to Monin-Obukhov similarity theory of the turbulent boundary layer flow, such as $u(z^+) = (1/\kappa)\ln z^+ + C^+$, every  surface exhibits some effects of roughness, so that the constant $C^+$ becomes a function of the roughness length $z_0^+$ such that $\lim_{z_0^+\rightarrow 0}C^+(z_0^+) = 5.0$. Here, $\kappa$ is the von Karman constant, and $z^+ = u_*z/\nu$ is the dimensionless distance from the wall, where $u_*$ is the wall-friction velocity, $z$ measures the wall-normal distance, and $\nu$ is the kinematic viscosity. Fig~\ref{lesdns} shows that the proposed near-surface model adequately captures the turbulent channel flow.
\subsection{Comparison with reference data: wind tunnel and LES}
\begin{figure}[H]
\emph{a)}	\centering{\psfig{figure=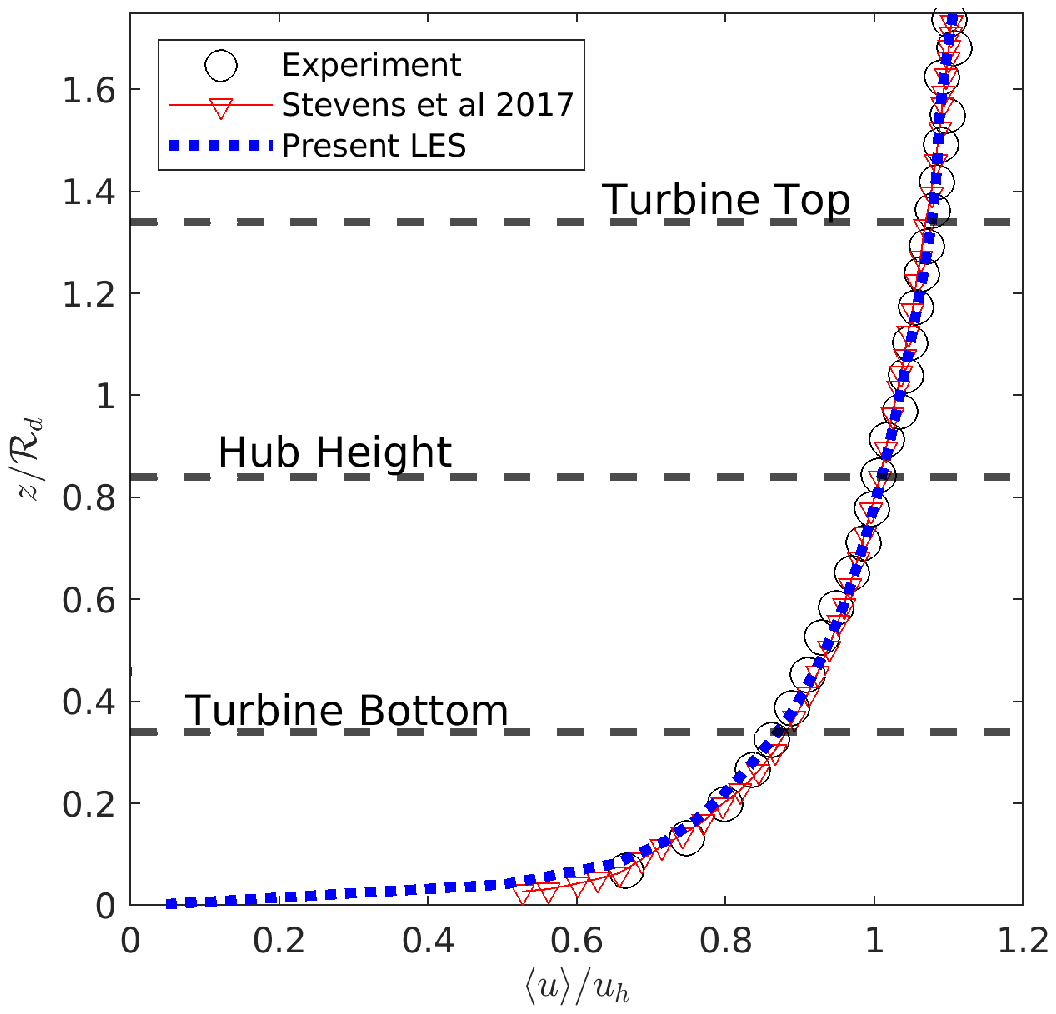,width=0.28\columnwidth}}
\emph{b)}	\centering{\psfig{figure=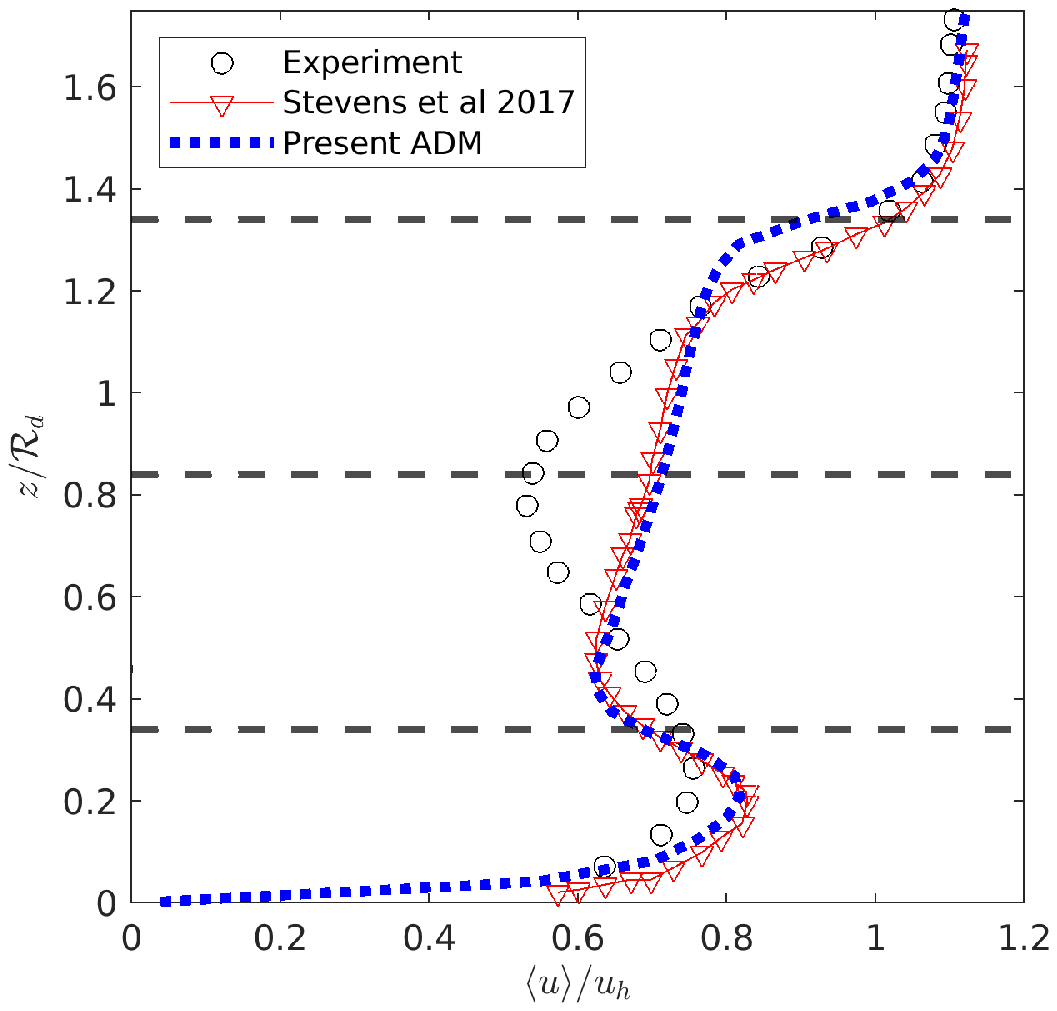,width=0.28\columnwidth}}
\emph{c)}	\centering{\psfig{figure=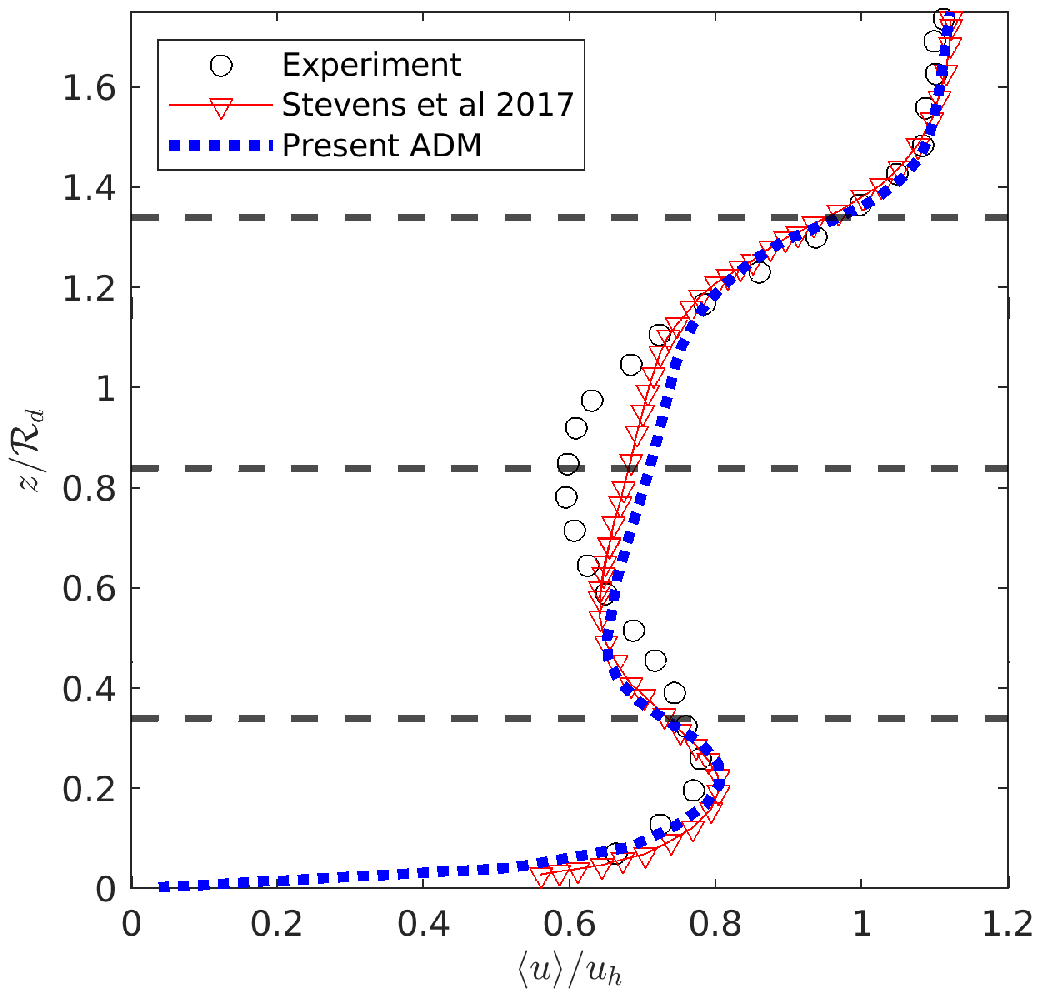,width=0.28\columnwidth}}
	\caption{Comparison of mean streamwise velocity at $-1 \mathcal{R}_{d}$, $2 \mathcal{R}_{d}$ and $3 \mathcal{R}_{d}$ respectively of present ADM model with experimental data \cite{porte2011large} and LES data \cite{stevens2018comparison}.}
	\label{singleturbine}
\end{figure}

In LES of atmospheric turbulence over wind farms, resolving the blades is computationally expensive. We consider two dataset to validate the actuator disk model for single wind turbine. We obtained the data from the wind-tunnel experiment conducted by \cite{porte2011large} and the LES data for actuator disk model adapted from \cite{stevens2018comparison}. Based on the experiment and previous numerical studies, we consider a computational domain of $\left[5.4 \times 0.9 \times 0.45\right]~m^{3}$, which is discretized into $256\times32\times72$ finite volume cells. In this study, we use the wind turbine model based on \cite{porte2011large} with rotor diameter ($\mathcal{R}_{d}$) of $0.15$~m, hub height ($\mathcal{H}_{hub}$) of $0.125$~m and coefficient of thrust, $C_t = 0.5669$ ($c_{t}=0.8248$). Considering the actuator disk approach, we find that  the wind turbine wake flow is accurately predicted when the proposed near-surface model  is considered. In Fig~\ref{singleturbine}, we observe that there is a good agreement among the present LES and the corresponding LES results~\cite{stevens2018comparison}. Present LES also shows an average agreement with the wind tunnel measurements\cite{porte2011large}.
\subsection{Interaction of wind farm with ABL}
Finally, we consider a LES study for a utility-scale wind farm, the layout of which is depicted in Fig~\ref{layout}. In this simulation, the horizontal grid spacing is approximately $11$~m, while the rotor diameter is $126$~m. Clearly, such a grid spacing is insufficient to resolve the tip vortices generated by turbines.
This section represents how the ABL responds to an array of wind turbines which operate on a fully aerodynamic rough surface. The shear stress exerted by the ABL flow over an aerodynamic rough surface is dominated by the Reynolds stress. In particular the flow near a rough surface accelerates considerably before being retarded by the influence of the turbines. It is necessary that the inlet profiles, the ground shear stress, and the turbulence model should be in equilibrium in close proximity to a rough surface. At the inflow boundary (e.g. $x = -1000$ in Fig~\ref{layout}), we have imposed the undisturbed mean logarithmic profile of ABL flow for the streamwise component of the velocity. On the inlet plane of the computational domain, we have introduced $m$ inviscid large eddies. The strength and intensity of such eddies are chosen randomly in the current study; however, the model is designed in such a way that the strength of such eddies in the inlet plane can be derived from an accompanying meteorological simulation or field measurements, as appropriate.

\begin{figure}[H]
	\centering{\psfig{figure=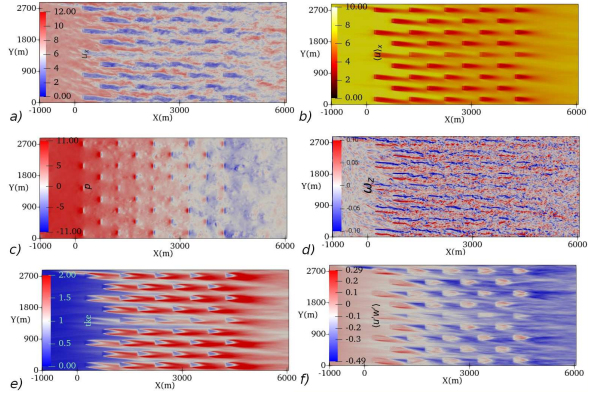,width=0.9\columnwidth}}
	\caption{Color filled contour plots based on proposed near-surface model in conjunction with square of the velocity gradient sgs model on a $x-z$ plane cutting through the hub height $\mathcal{H}_{hub}=90$~m of the turbines a) $u_x$, instantenous stream-wise velocity b) $\langle u_{x} \rangle$, mean stream-velocity c) Pressure field,  d) $\omega_{z}$, vorticity in the surface normal direction e) TKE, turbulent kinetic energy and f) $\langle u'w' \rangle$, vertical flux.}%
	\label{colorplot}
\end{figure}

\begin{figure}[H]
\emph{a)} \centering{\psfig{figure=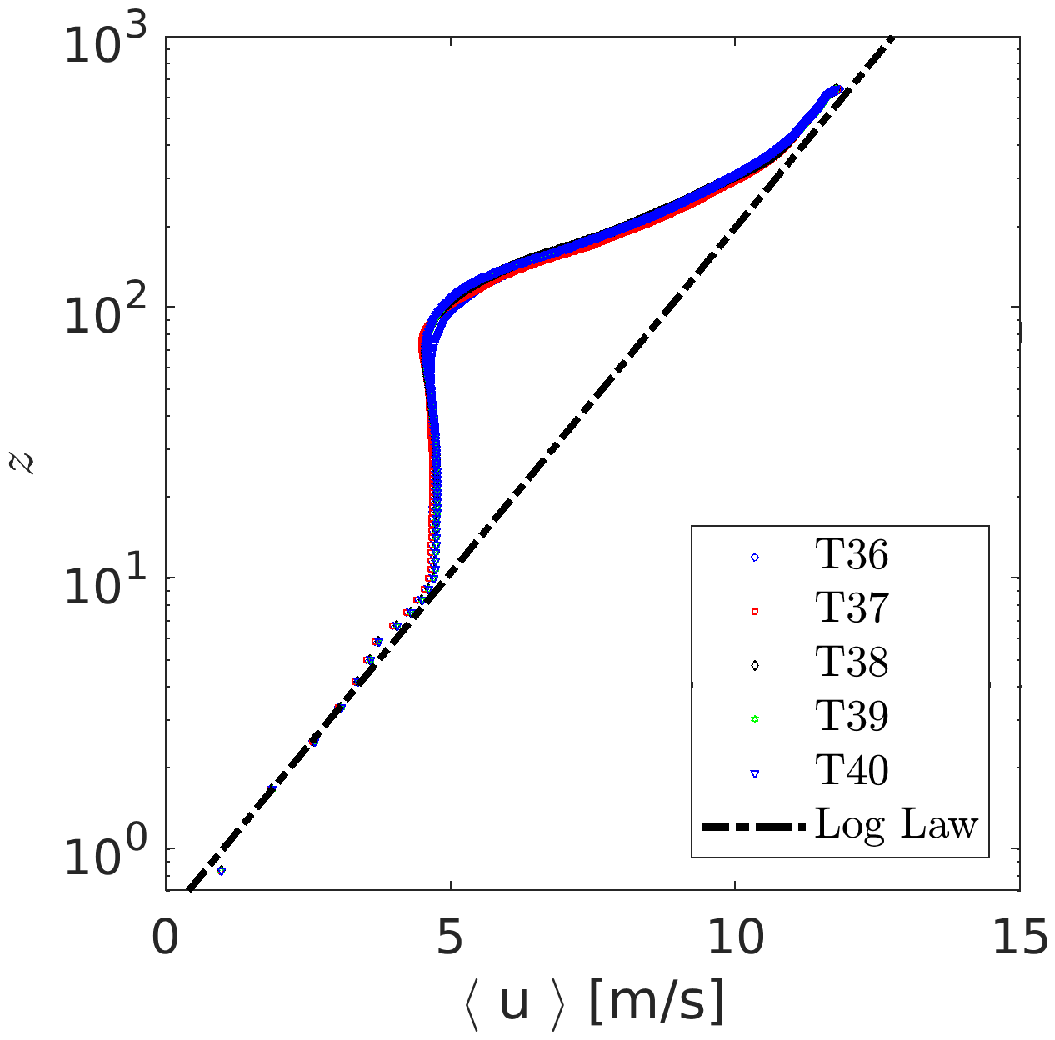,width=0.4\columnwidth}}
\emph{b)}	\centering{\psfig{figure=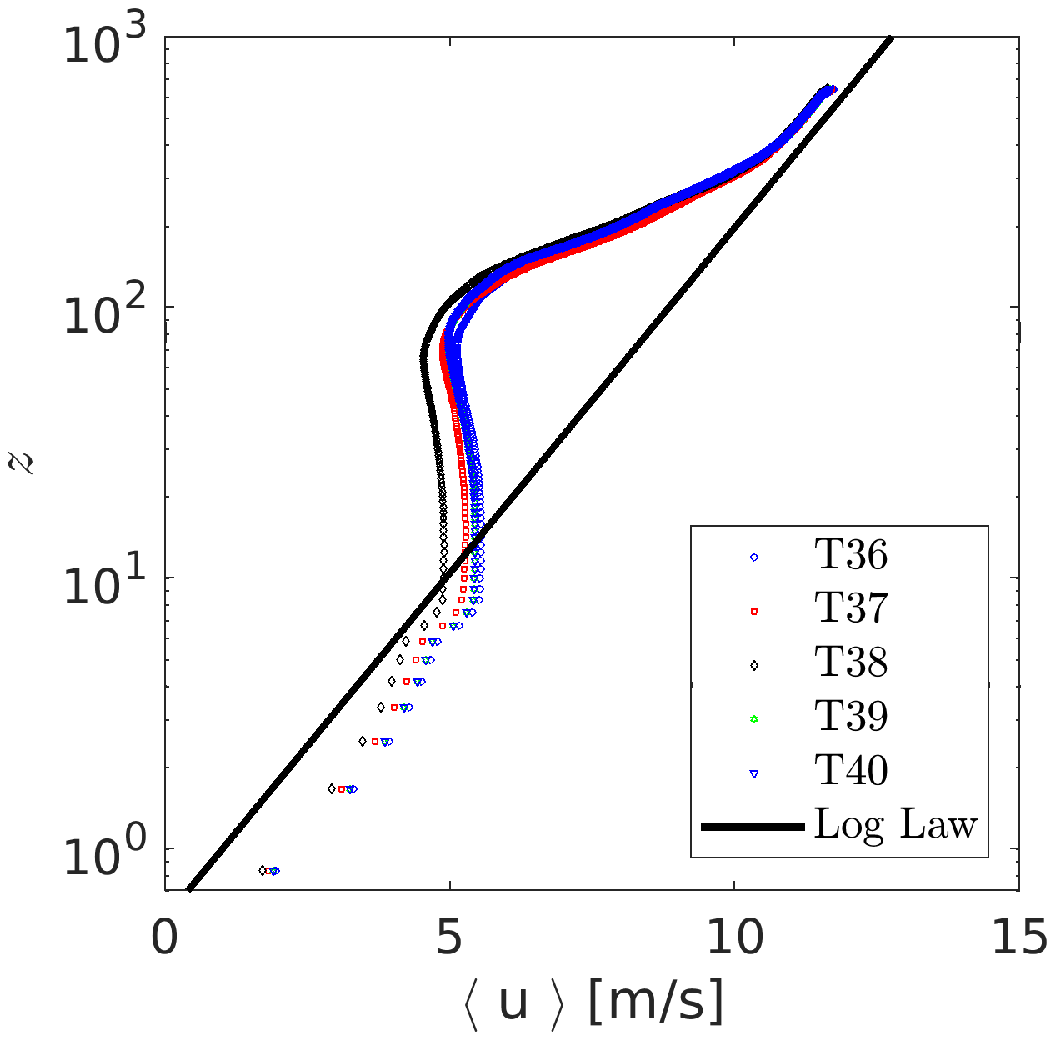,width=0.4\columnwidth}}
	\caption{A comparison of velocity deficit behind the wind turbines in utility-scale wind farm. (a) proposed near-surface model. (b) classical wall-stress model for rough-surface ABL \cite{moeng1984large}.}
	\label{ns}
\end{figure}

Fig \ref{colorplot} display the color-filled contour plots of the streamwise instantaneous velocity (Fig \ref{colorplot}a), the mean streamwise velocity (Fig \ref{colorplot}b), and pressure field (Fig \ref{colorplot}c). Fig \ref{colorplot}d display the
vertical component of vorticity, resolved turbulence kinetic energy (Fig \ref{colorplot}e), and resolved fraction of the Reynolds stress (Fig \ref{colorplot}f). These color-filled contour plots illustrates the instantaneous flow on the horizontal plane passing through the hub-height $z = 90$~m. Fig \ref{colorplot}(a--b) indicates that the first two rows are exposed directly to the wind that was imposed numerically at the inlet boundary $x = -1000$. The asymptotic velocity deficit in rows 2-4 are influenced by the wakes from upstream turbines. The boundary layer is fully developed in the region downstream to row 5. It is worth mentioning that the number of turbines (e.g. 41) and the size of the domain is limited by the computational resources and may not be sufficient to reach a fully developed `large wind-farm'. Such limitations are to be considered while interpreting our LES data. Nevertheless, the overall flow pattern is consistent with the expected regime of turbulent flow over fully developed wind farms.
Fig~\ref{ns}a shows that the vertical profile of the velocity agrees with the logarithmic profile in regions below and above the wind turbine array. Moreover, the velocity deficit in Fig~\ref{ns} exhibits a close agreement with the corresponding profile of wind tunnel measurement depicted in Fig~\ref{singleturbine}b. Fig~\ref{ns}b shows the profile of velocity deficit, where the classical near-wall model is employed.

\begin{figure}[H]
\emph{a)}	\centering{\psfig{figure=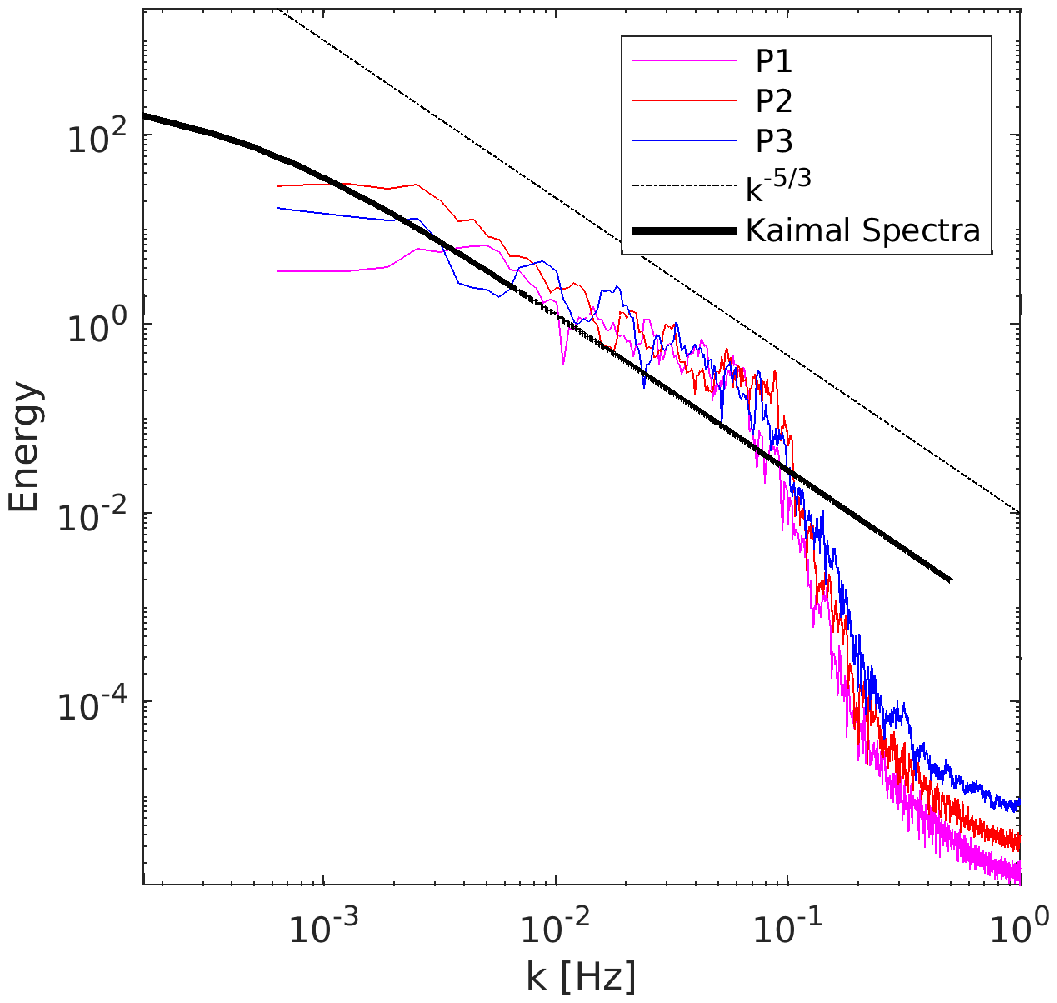,width=0.4\columnwidth}}
\emph{b)}	\centering{\psfig{figure=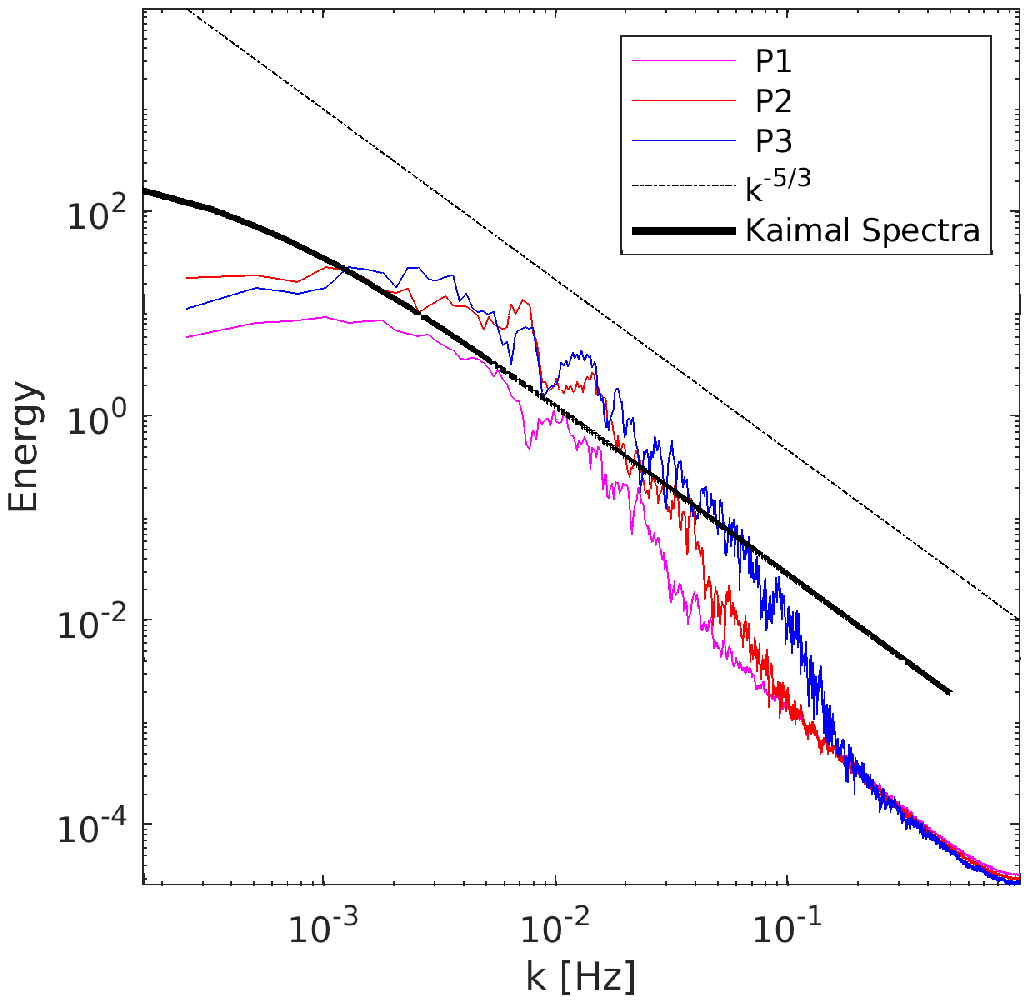,width=0.4\columnwidth}}
	\caption{Energy spectra of streamwise velocity captured at P1($716, 1440, 27$),P2($2732, 1440, 27$) and P3($4244, 1440, 27$). a) Proposed near-surface model in conjunction with square of the velocity gradient sgs model. b) A subgrid model commonly used in ABL flow simulation  \cite[e.g., TKE-1.5 model,][]{deardorff1972numerical}.}
	\label{spectra}
\end{figure}

\begin{figure}[H]
\emph{a)}	\centering{\psfig{figure=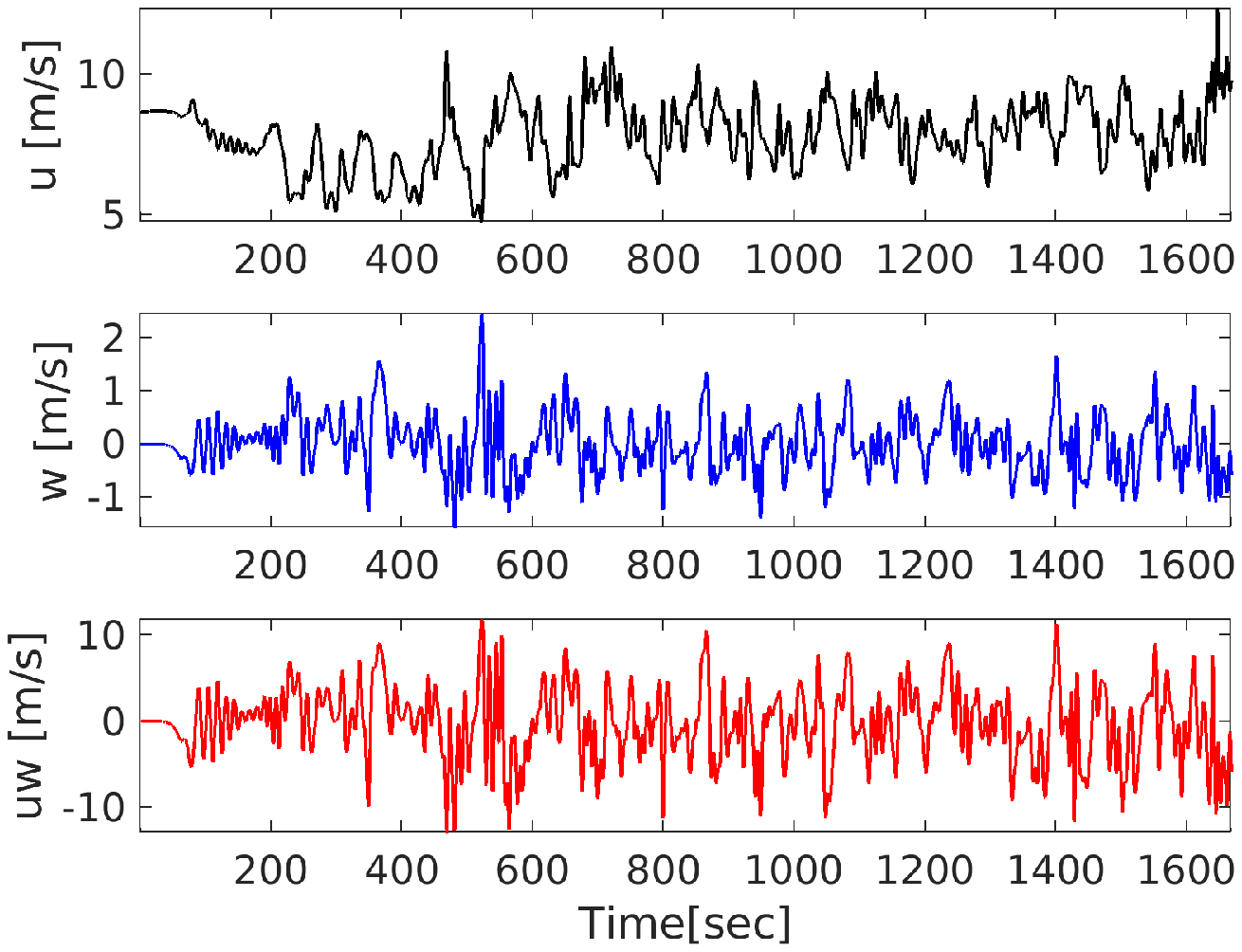,width=0.4\columnwidth}}
\emph{b)}	\centering{\psfig{figure=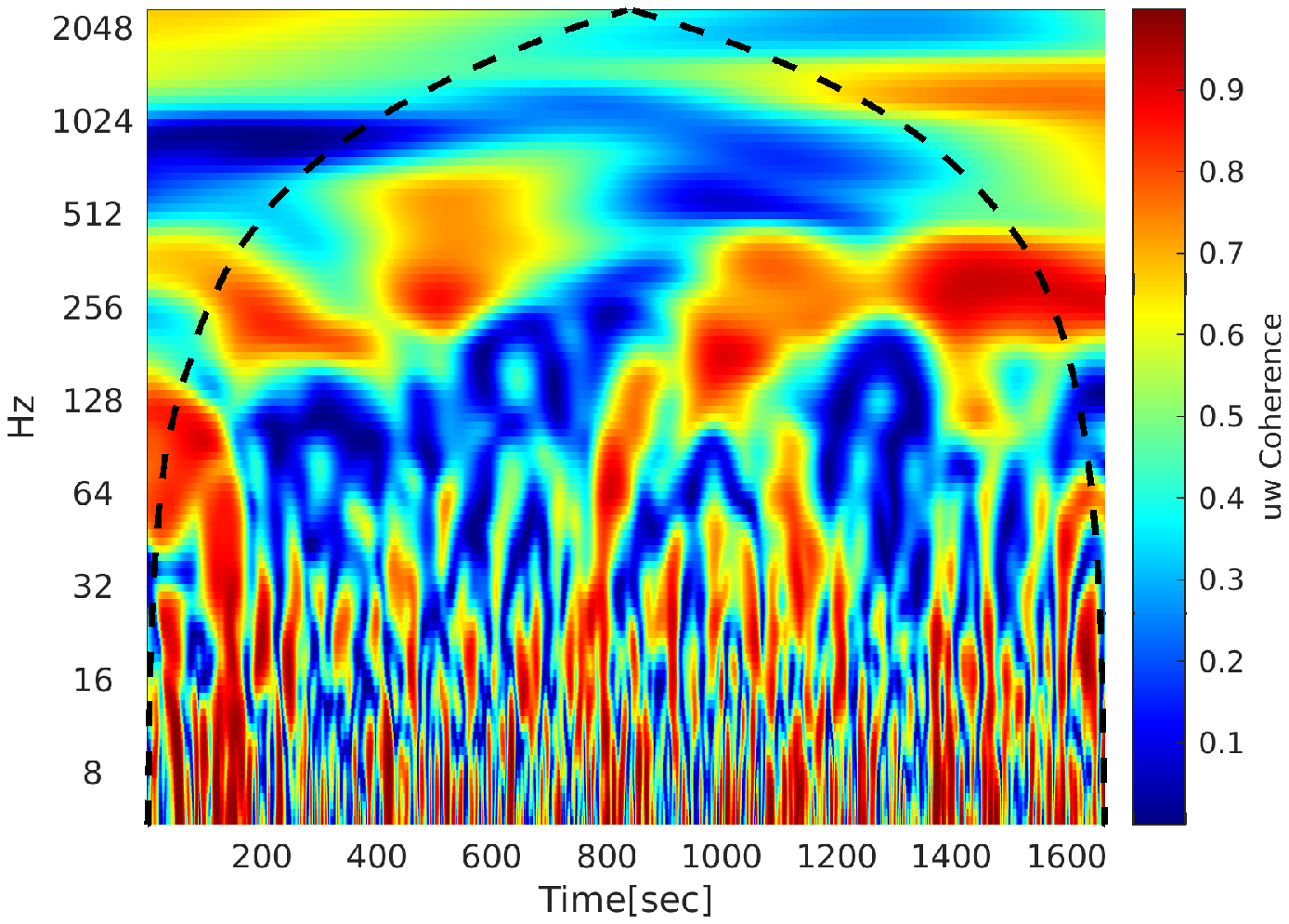,width=0.415\columnwidth}}
	\caption{a) Instantaneous vertical flux.~b) wavelet power spectral density of the vertical flux captured at P1$(716,1440,163)$.} 
	\label{coherence}
\end{figure}

In order to understand the influence of inflow turbulence, Fig~\ref{spectra} compares the energy spectrum with respect to the two turbulence models. A theoretical -5/3 slope proposed by Kolmogorov and Kaimal spectra  is considered to verify the capability of  LES model  to reproduce the energy cascade. 
In Fig~\ref{coherence}, we consider the wavelet coherency diagram for the wind around the turbine \#1 (see layout in Fig~\ref{layout}). We employed the wavelet transform technique to understand the role of coherent structures in the wind farm. In comparison to fourier transform, wavelet transform can extract both local spectrum and temporal details. It can be clearly seen from the Fig \ref{coherence}b that energy spectrum is associated with the transient coherent vorticies.
\section{Conclusion}
We have developed a dynamic procedure of modeling sub-
grid scale near-surface atmospheric turbulence in large wind
farms. This approach resolves neither the near-surface region nor the wind turbine blades. A salient feature is the
dynamic blending of shear stress, while implementing the
Dirichlet boundary conditions. In this model the scaling of
the eddy viscosity is achieved dynamically, which is based
on the vortex stretching vector, instead of the classical scal-
ing $\Delta^{2}$, where $\Delta$ is the grid spacing. Unlike the classical dynamic approach of modelling subgrid scale stress, the fraction of the desired level of dissipation is fixed through a global value of the model parameter $C_{w}$. However, the exchange of momentum between the earth's surface and the atmosphere aloft assumes a local dynamic variation of Monin-Obukhov similarity profile at each grid point adjacent to the surface.

In comparison to classical wall-stress model for rough wall
atmospheric turbulence combined with a kinetic energy based
subgrid model (commonly utilized in atmospheric boundary layer studies), the results indicate that the new model is capable to accurately predict both the wind turbine wake dynamics and the near-surface atmospheric turbulence. It is also observed that the energy spectrum follows the classical power law $k^{-5/3}$
in case of the proposed model, exhibiting a `sharp
spectral cutoff' although we have considered a finite-volume
discretization. Using wavelet coherency diagram, we observe
that the energy spectrum is indeed associated with the transient coherent vortices.

We plan to discuss further advancement of this development
in another article, which is currently in progress. In particular, the underlying symmetries of the differential operators are
preserved in the numerical scheme that we have implemented
within our in-house LES code. Although the actuator disk
approach is considered, we are working on improved modelling of the thrust of wind turbines through the advancement of the canopy-stress model in the individual wake of turbines in large wind farms. Canopy-stress model is considered to be an effective numerical tool to model ABL, as well as the effects of mountain-like obstacles in the atmospheric boundary layer.  

\bibliography{ref.bib}

\end{multicols}

\end{document}